\documentclass[aps,prl,superscriptaddress,12pt]{revtex4}

\usepackage{times}

\usepackage{amsmath}
\usepackage{amsfonts}
\usepackage{amssymb}
\usepackage{epsfig}

\begin{document}

\title{Coherence and Indistinguishability of Single Electrons Emitted by Independent Sources}

\author
{E. Bocquillon,$^{1}$ V. Freulon,$^{1}$ J.-M Berroir,$^{1}$ P. Degiovanni,$^{2}$\\
B. Pla\c{c}ais,$^{1}$ A. Cavanna,$^{3}$ Y. Jin,$^{3}$ G. F{\`e}ve$^{1\ast}$ \\
\normalsize{$^{1}$Laboratoire Pierre Aigrain, Ecole Normale Sup\'erieure, CNRS (UMR8551), Universit\'e Pierre et Marie Curie, Universit\'e Paris Diderot}\\
\normalsize{24 rue Lhomond, 75231 Paris Cedex
05, France}\\
\normalsize{$^{2}$ Universit\'e de Lyon, F\'ed\'eration de Physique Andr\'e Marie Amp\`ere,} \\
\normalsize{CNRS - Laboratoire de Physique de l'Ecole Normale Sup\'erieure de Lyon} \\
\normalsize{46 All\'ee d'Italie, 69364 Lyon Cedex 07,France.}\\
\normalsize{$^{3}$CNRS - Laboratoire de Photonique et de Nanostructures}\\
\normalsize{Route de Nozay, 91460 Marcoussis, France}\\
\normalsize{$^\ast$ To whom correspondence should be addressed;
E-mail:  feve@lpa.ens.fr.} }

\baselineskip24pt


\begin{abstract}
The on-demand emission of coherent and indistinguishable electrons by independent synchronized sources is a challenging task of quantum electronics, in particular regarding its application for quantum information processing. Using two independent on-demand electron sources, we trigger the emission of two single-electron wavepackets at different inputs of an electronic beamsplitter. Whereas classical particles would be randomly partitioned by the splitter, we observe two-particle interferences resulting from quantum exchange. Both electrons, emitted in indistinguishable wavepackets with synchronized arrival time on the splitter, exit in different outputs as recorded by the low frequency current noise. The demonstration of two-electron interference provides the possibility to manipulate coherent and indistinguishable single-electron wavepackets in quantum conductors.
\end{abstract}
\date{\today}
\maketitle

As for photons, the wave-particle duality plays a crucial role in the propagation of electrons in quantum conductors. The wave nature of electrons can be revealed in interference experiments ({\it 1--3}) probing the single-particle coherence of electron sources through the measurement of the average electrical current. The corpuscular nature of charge carriers shows up when measuring fluctuations of the electrical current ({\it 4}).
Still, a few experiments cannot be understood within the wave nor the corpuscular description: this is the case when two-particle interferences effects related to the exchange between two indistinguishable particles take place. These experiments have proven particularly interesting, first on a fundamental point of view as they require a full quantum treatment, and secondly, because the on-demand generation of indistinguishable partners is at the heart of quantum information protocols ({\it 5}). Information coding in few electron states that propagate ballistically in quantum conductors ({\it 6}) thus requires the production of coherent and indistinguishable single-particle wavepackets emitted by several synchronized but otherwise independent  emitters. The collision of two particles emitted at two different inputs of a beamsplitter can be used to measure their degree of indistinguishability.  In the case of bosons, indistinguishable partners always exit in the same output (see Fig. 1). Fermionic statistics leads to the opposite behavior: particles exit in different outputs. The bunching of indistinguishable photons has been observed by recording the coincidence counts between two detectors placed at the outputs of the beamsplitter as a function of the time delay $\tau$ between the arrival times of the photons on the splitter. Bunching shows up in a dip in the coincidence counts, the Hong-Ou-Mandel (HOM) dip ({\it 7}), when $\tau$ is varied. The reduction of the coincidence counts directly measures the overlap between the single-particle states at the input. It is maximum when the arrivals are synchronized and can be suppressed when the delay becomes larger than the wavepacket widths.

The production of indistinguishable partners is challenging and their generation by independent sources has been only recently achieved in optics  ({\it 8}). In one dimensional quantum conductors, a continuous stream of indistinguishable electrons can be produced by applying a dc voltage bias to two different electronic reservoirs. Due to fermionic statistics, each source fills the electronic states up to the chemical potential $-eV$ and identical electron beams are generated. Using such sources, the $\pi$ exchange phase of indistinguishable fermions has been measured in the above described collider geometry ({\it 9}) and in a two-particle interferometer based on a Mach-Zehnder geometry ({\it 10, 11}).  However, as these sources generate a continuous beam of electrons, they do not reach the single particle resolution of their optical analog and two-particle interferences cannot be interpreted as resulting from the overlap between two single particle wavepackets. The manipulation of single-particle states thus requires to replace dc emitters by triggered ac emitters that generate a single-electron wavepacket at a well defined time.

Dealing with electrons, one can benefit from the charge quantization of a small quantum dot enforced both by Coulomb interaction and fermionic statistics to trigger the emission of particles one by one ({\it 12--16}). Moreover, the edge channels of the quantum Hall effect provide an ideal test bench to implement optic-like experiments with electron beams in condensed matter, as electron propagation is ballistic, one-dimensional and chiral. We will consider here a mesoscopic capacitor ({\it 12}), which comprises a small quantum dot capacitively coupled to a metallic top gate and tunnel coupled to a single edge channel by a quantum point contact of variable transmission $D$. By applying a square wave periodic rf excitation on the top gate which peak to peak amplitude matches the dot addition energy, $2 e V_{exc} \approx \Delta$, a quantized current resulting from the emission of a single electron followed by a single hole is generated ({\it 12}). Beyond average current measurements, this emitter has been characterized through the study of current correlations on short times ({\it 17--20}) as well as partition noise measurements ({\it 21}) in the electronic analog of the Hanbury-Brown and Twiss geometry ({\it 22, 23}). These measurements have demonstrated that, for escape times smaller than half the period of the excitation drive,  exactly a single-electron and a single-hole excitations were successively emitted at each period. Moreover, the tunnel emission of single particles from a discrete dot level should lead to electron and hole wavefunctions described by exponentially decaying wavepackets ({\it 24, 25}): $\phi (t) =\frac{1}{\sqrt{\tau_e}} \Theta(t-t_0) \; e^{i \frac{\Delta (t-t_0)}{2 \hbar} } \; e^{- \frac{t-t_0}{2 \tau_e }}$, where $\Theta(t)$ is the step function, $\Delta/2$ is the energy of emitted electrons and holes, and $t_0$ is the emission trigger that can be tuned with a few picoseconds accuracy. Measurements of the average current $\langle I(t) \rangle $ ({\it 12}) and short-time correlations  $\langle I(t) I(t+\tau) \rangle$ ({\it 17}) have confirmed that the probability of single-particle detection (that is the envelope of the wavepacket) was following this exponential decay. However, these measurements are only sensitive to the squared modulus of the wavefunction, $|\phi(t)|^2$ and as such, do not probe the coherence of the electronic wavepacket related to the phase relationship between $\phi(t)$ and $\phi^{*}(t')$  (for $t \neq t'$) and encoded in the off-diagonal components (coherences) of the density matrix $\rho(t,t') = \phi (t) \phi^{*}(t')$.

Using two such emitters at the two inputs of an electronic beamsplitter, the coherence and indistinguishability of two single electronic wavepackets can be probed by two-electron interferences ({\it 25--27}). Considering the electron emission sequence, each emitter generates an electronic wavepacket $|\phi_i \rangle $ ($i=1,2$) above the Fermi energy at each input of the splitter set at transmission $\mathcal{T}=1/2$. The probability $P(1,1)$ that the two particles exit in different outputs is related to the overlap between wavepackets: $P(1,1) = \frac{1}{2} \left[ 1 + |\langle \phi_1 | \phi_2\rangle |^2 \right]$. An opposite sign occurs in the expression of the probability that both particles exit in the same output, $P(0,2) + P(2,0) = \frac{1}{2} \left[ 1 - |\langle \phi_1 | \phi_2\rangle |^2 \right]$. These signs are related to the exchange phase of $\pi$ for fermions, they would be opposite for bosons. For fermions, the coincidence counts for indistinguishable particles would thus be doubled compared to the classical case (Fig. 1). However, single shot detection of ballistic electrons in condensed matter is not available. Antibunching is thus not probed by coincidence counts but rather by low frequency fluctuations of the electrical current in the outputs related to the fluctuations of the number of transmitted particles: $\langle \delta N_3 ^2 \rangle = \langle \delta N_4 ^2 \rangle = \frac{1}{2} \left[ 1 - |\langle \phi_1 | \phi_2\rangle |^2 \right] $. Repeating this two-electron collision at frequency $f$, and considering the successive emission of one electron and one hole per period, the low frequency current noise at the output is then given by ({\it 25}):
\begin{eqnarray}
S_{33} & =& S_{44} = e^2 f \times \left[1 - \left|\langle \phi_1 | \phi_2\rangle \right| ^2 \right]\\
& =&  e^2 f \times \left[1 - \left| \int dt \; \phi_1(t) \phi_{2}^{*}(t) \right|^2 \right]
\end{eqnarray}
Note that the single-electron wavepackets $\phi_i$ in Eq.(2) differ from the states generated by applying a time-dependent voltage $V_{i}(t)$ on each electronic reservoir connected to inputs $i=1,2$ and cannot be generated by such classical drive (in which case, the two inputs in Eq.(2) can be reduced to a single one by the proper gauge transformation that shifts the potentials by $V(t)=V_2(t)$).
For perfectly indistinguishable states, $\phi_2(t) = \phi_1(t)$, a complete suppression of the output noise is obtained. By delaying by time $\tau$ the emission of one particle with respect to the other: $\phi_2(t) = \phi_1 (t + \tau)$, the full random partitioning of classical particles $S_{33}  = S_{44} = e^2 f$ can be recovered (Fig. 1). It is thus convenient to consider the noise normalized by the classical random partitioning $q=S_{44}/e^2f$ which equals for exponentially decaying wavepackets:
\begin{eqnarray}
q & = & 1-e^{-|\tau|/\tau_e}
\end{eqnarray}
Note that Eq.(3) is valid at zero temperature, or when the Fourier components of the wavefunctions $\tilde{\phi_i}(\omega)$ have no overlap with the thermal excitations: $\tilde{\phi_i}(\omega) = 0$ for $\hbar \omega \approx k_B T$. Otherwise, the random partitioning is also affected by antibunching with the thermal excitations, so that $S_{44} \leq e^2 f$ ({\it 21}). However, if one measures the normalized value of the excess noise $\Delta q$, between the situations where both sources are switched on and switched off, simulations describing the source in the Floquet scattering formalism ({\it 20, 28}) show that $\Delta q$ is accurately described by Eq.(3) for moderate temperatures $k_B T \ll \Delta$.

The circuit (Fig. 2), is realized in a 2D electron gas
(2DEG) at a AlGaAs/GaAs heterojunction, of nominal density
$n_{s}=1.9 \times 10^{15}{\rm\, m^{-2}}$ and mobility $\mu =2.4 \times 10^6\, {\rm cm^{2} V^{-1} s^{-1}}$. A strong magnetic field $B=2.68$ T is applied so as to work in the quantum Hall regime at filling factor $\nu=3$  ($\nu=3$ is chosen because, in this sample, the splitter transparency $\mathcal{T}$ becomes energy dependent at higher values of the magnetic field). Two mesoscopic capacitors with identical addition energies $\Delta = 1.4$ K (much larger than the electronic temperature $T=100$ mK) are used as electron/hole emitters and placed at a $3\ \mu$m distance from a quantum point contact used as an electronic beamsplitter at transmission $\mathcal{T}=\frac{1}{2}$. Single charge emission in the outer edge channel is triggered with a square excitation at frequency $f=2.1$ GHz with average emission times set to $\tau_{e,1}= \tau_{e,2} =58 \pm 7$ ps  corresponding to a transmission $D_1=D_2=0.45 \pm 0.05$. The low frequency partition noise is measured at output 3.  Fig. 3 presents the measurements of $\Delta q$ as a function of the time delay $\tau$ between the two sources. We observe a dip in the noise measurements for zero time delay and a plateau for longer time delays. The noise values $\Delta q$ are normalized by the value of the noise on the plateau. The
sum of the partition noises for each source can also be measured by switching off each source
alternately. This random partition noise is represented on Fig. 3 by the blurry blue line, which
extension represents the error bar. As expected, it agrees with $\Delta q$ for large time delays.

The dip observed for short time delay is analogous to the HOM dip but is related here to the antibunching of single indistinguishable fermions, we thus call it the Pauli dip. It  reflects our ability to produce single-particle states emitted by two different emitters with some degree of indistinguishability. The states are not perfectly identical as shown by the fact that the dip does not go to zero. Note that Coulomb repulsion between electrons and between holes on the splitter could also be responsible for a dip in the low frequency noise. However, this effect can be ruled out using the long time delay limit, $\tau \approx 240$ ps. In this limit, the arrival of one electron is synchronized with the arrival of a hole in the other input. As can be seen on Fig. 3, a random partitioning is observed while Coulomb attraction between electron and holes would also predict a dip in the low frequency noise (as the transmitted charge is always zero when electrons and holes exit in the same output). The dip around zero time delay can be well fitted by the expression $\Delta q = 1 - \gamma e^{- \frac{\tau -\tau_0}{\tau_e}}$ expected for two exponentially decaying wavepackets but with a non unit overlap $\gamma$. We find $\tau_e = 62 \pm 10$ ps, $\gamma = 0.45 \pm 0.05$ and $\tau_0 = 13 \pm 6$ ps, consistent with the ten picoseconds accuracy of the synchronization between sources. As mentioned above, these results can be compared with a numerical simulation of $\Delta q$ in the Floquet scattering formalism, which we denote $\Delta q_F(\tau)$. For identical emission parameters of both sources, Floquet theory predicts a unit overlap at zero time delay, $\Delta q_F(\tau=0)=0$. The red trace on Fig. 3 represents $\Delta q= 1- \gamma (1-\Delta q_F (\tau))$ which imposes a non unit overlap $\gamma $ in the Floquet scattering formalism. It reproduces well the shape of the dip using the following parameters: $\gamma=0.5$,  $D_1=D_2=0.4$, $\Delta_1= \Delta_2 = 1.4 \ \rm K$ and $T=100 \ \rm mK$.

This non unit overlap can be attributed to two different origins. First, it could stem from some small differences in the emission energies related to small differences in the static potential of each dot. Using Eq.(2), a reduction to a $50 \%$ overlap can be obtained by shifting one level compared to the other by energy $\Delta/10$. The value of the static potential is fixed with a better accuracy but small variations could occur within the several hours of measurement time for each point. The second possibility is related to the decoherence of single-electron wavepackets during propagation towards the splitter (that could arise from Coulomb interaction with the adjacent edge channel). In a simple treatment of the wavepacket decoherence, the pure state $\phi_1(t)$ is replaced by the density matrix $\rho_1(t,t') = \phi_1(t) \phi_1^{*}(t') \mathcal{D}_{1}(t,t') $ where $\mathcal{D}_1(t,t')$ is a decoherence factor ({\it 27, 29}). We have $\mathcal{D}_1(t,t)=1$, such that the average current $\langle I(t) \rangle$ is not affected, but $\mathcal{D}_1(t,t') \rightarrow 0$ for $|t-t'| \rightarrow \infty$, suppressing the coherence of the electronic wavepacket. In that case, Eq.(2) becomes:
\begin{eqnarray}
\Delta q & = &  1 - {\rm Tr}[\rho_1 \rho_2] \\
 & =& 1 - \int \ dt \ dt'\ \phi_1(t) \phi_1^{*}(t') \mathcal{D}_{1}(t,t') \phi_2^{*}(t) \phi_2(t') \mathcal{D}_{2}(t,t')
\end{eqnarray}
Eq.(5) exemplifies the fact that the noise suppression stems from a two-particle interference effect encoded in the off-diagonal components of the density matrices $\rho_i$, i.e. on the coherence of the electronic wavepacket.  Assuming $\mathcal{D}_1(t,t') = \mathcal{D}_{2}(t,t') = e^{-\frac{|t-t'|}{\tau_{c}}}$  in Eq.(5), we find analytically that the overlap depends on the ratio between the intrinsic coherence time of the wavepacket $\tau_e$ and the coherence time $\tau_{c}$ associated with the propagation along the edge: $\gamma = \frac{\tau_{c}/(2 \tau_e)}{1 + \tau_{c}/(2 \tau_e)}$. For $\tau_e \ll \tau_{c}$, the effects of decoherence can be neglected but in the opposite limit, $\tau_{c} \ll \tau_{e}$, the overlap is completely suppressed and the classical partitioning is recovered. In this case, electrons are rendered distinguishable through their interaction with the environment. Within this picture, our measurement of the overlap is compatible with $\tau_{c} \approx 100 \ \rm ps$. Such decoherence effects underline the necessity to reach the subnanosecond timescale in electron emission to be able to generate indistinguishable electron wavepackets.

The observed Pauli dip in the low frequency noise of the output current for short time delays between the arrival times of electrons at a beam splitter is a signature of two-particle interferences which demonstrates the possibility to generate coherent and indistinguishable single-electron wavepackets with independent sources. It provides the possibility of controlled manipulation of single-electron states in quantum conductors, with applications in quantum information processing, but could also be used to fully reconstruct the wavefunction of a single electron ({\it 24, 30}) and thus quantitatively address the propagation of a single excitation propagating in a complex environment.

{\bf References and notes:}

\begin{enumerate}

\item
Y. Ji {\it et al.}, {\it  Nature} {\bf 422}, 415 (2003).

\item
P. Roulleau {\it et al.}, {\it Phys. Rev. Lett.} {\bf 100}, 126802 (2008).

\item
M. Yamamoto {\it et al.}, {\it Nature Nanotechnology} {\bf 7}, 247 (2012).

\item
Y. Blanter, and M. B\"{u}ttiker, {\it Physics Reports}  {\bf 336}, 1 (2000).

\item
E. Knill, R. Laflamme, and G.J. Milburn, {\it Nature} {\bf 409}, 46 (2001).

\item
A. Bertoni, P. Bordone, R. Brunetti, C. Jacoboni, and S. Reggiani, {\it Phys. Rev. Lett.} {\bf 84},  5912 (2000).

\item
C. K. Hong, Z. Y. Ou, and L. Mandel, {\it Phys. Rev. Lett.} {\bf 59}, 2044 (1987).

\item
J. Beugnon {\it et al.},  {\it Nature}  {\bf 440}, 779 (2006).

\item
R. C. Liu, B. Odom, Y. Yamamoto, and S. Tarucha, {\it Nature} {\bf 391},  263 (1997).

\item
 P. Samuelsson,  E. V. Sukhorukov and M. B\"{u}ttiker, {\it Phys. Rev. Lett.} {\bf 92}, 02685 (2004).

\item
I. Neder {\it et al.}, {\it Nature} {\bf 448}, 333 (2007).

\item
G. F\`{e}ve {\it et al.}, {\it Science}  {\bf 316}, 1169 (2007).

\item
 M. D. Blumenthal {\it et al.}, {\it Nature Phys.} {\bf 3}, 343 (2007).

\item
C. Leicht {\it et al.}, {\it Semicond. Sci. Technol.} {\bf 26}, 055010 (2011).

\item
S. Hermelin {\it et al.}, {\it Nature} {\bf 477}, 435 (2011).

\item
R. P. G. McNeil {\it et al.}, {\it Nature}  {\bf 477}, 439 (2011).

\item
A. Mah\'{e} {\it et al.}, {\it Phys. Rev. B} {\bf 82}, 201309 (R) (2010).

\item
 M. Albert, C. Flindt, and M. B\"uttiker, {\it Phys. Rev. B} {\bf 82}, 041407(R) (2010).

\item
  T. Jonckheere, T. Stoll, J. Rech, and T. Martin , {\it Phys. Rev. B} {\bf 85}, 045321 (2012).

\item
F.D. Parmentier {\it et al.},  {\it Physical Review B} {\bf 85}, 165438 (2012).

\item
E. Bocquillon {\it et al.}, {\it Phys. Rev. Lett.} {\bf 108}, 196803 (2012).

\item
M. Henny {\it et al.}, {\it Science} {\bf 284}, 296 (1999).

\item
W. Oliver, J. Kim, R. Liu, and Y. Yamamoto, {\it Science} {\bf 284}, 299 (1999).

\item
C. Grenier {\it et al.}, {\it New Journal of Physics} {\bf 13}, 093007 (2011).

\item
T. Jonckheere, J. Rech, C. Wahl, and T. Martin, {\it Phys. Rev. B}  {\bf 86},  125425 (2012).

\item
S. Ol'khovskaya, J. Splettstoesser, M. Moskalets, and M. B\"{u}ttiker, {\it Phys. Rev. Lett.} {\bf 101}, 166802 (2008).

\item
G. F\`{e}ve, P. Degiovanni and T. Jolicoeur, {\it Phys. Rev. {\bf B}}  {\bf 77}, 035308 (2008).

\item
M. Moskalets, P. Samuelsson, and M. B{\"u}ttiker,  {\it Phys. Rev. Lett.} {\bf 100}, 086601 (2008).

\item
P. Degiovanni, C. Grenier, and G. F\`{e}ve, {\it Phys. Rev. B} {\bf 80}, 241307 (R) (2009).

\item
G. Haack, M. Moskalets, J.Splettstoesser, and M. B{\"u}ttiker,  {\it Phys. Rev. B} {\bf 84}, 081303 (2011).

\end{enumerate}

{\bf Acknowledgments:}  This work is supported by the ANR grant '1shot',
ANR-2010-BLANC-0412.

\clearpage
\begin{figure}[hhhhhh]
\centerline{\includegraphics[width=17 cm, keepaspectratio]{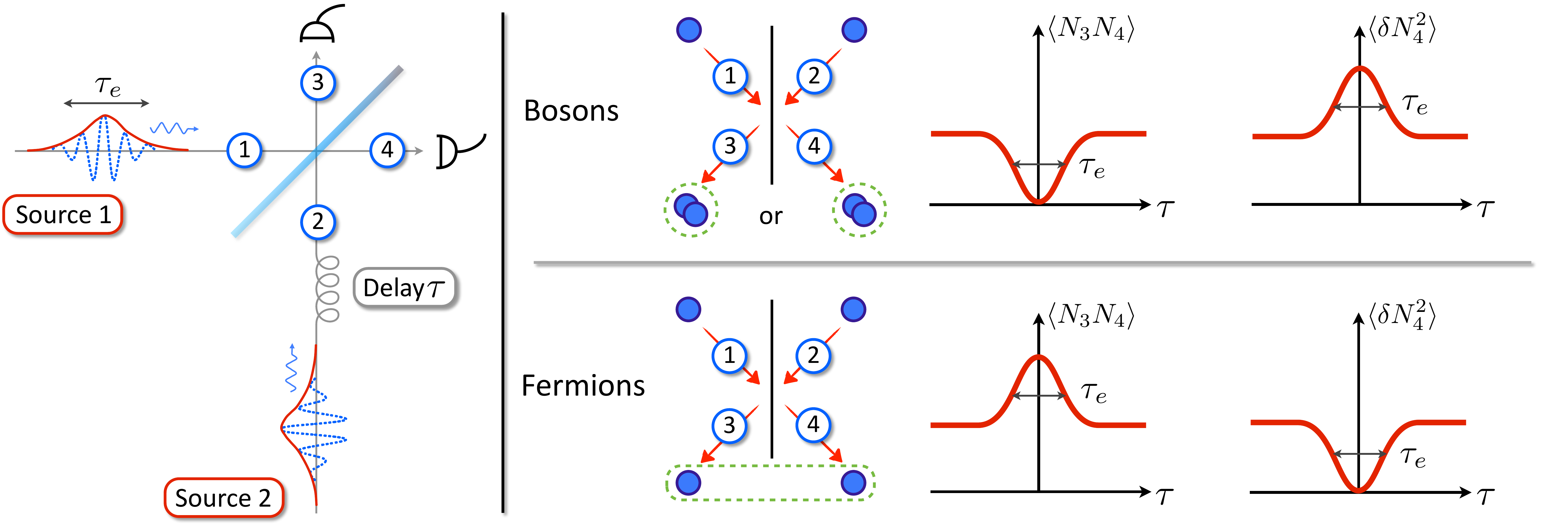}}
 \caption{Sketch of the experiment. Two single-particle wavepackets of width $\tau_e$ are emitted at inputs $1$ and $2$ and partitioned on a splitter. Coincident counts $\langle N_3 N_4 \rangle$ and fluctuations $ \langle \delta N_4 ^2 \rangle$ can be recorded at the outputs $3$ and $4$ as a function of the tunable delay $\tau$. Indistinguishable bosons always exit in the same output. This results in a suppression of the coincidence count and a doubling of the fluctuations at zero delay compared to the partitioning of classical particles obtained for $\tau \gg \tau_e$. An opposite behavior is expected for indistinguishable fermions (doubling of the coincidence counts and suppression of the fluctuations).}
\end{figure}

\begin{figure}[hhhhhh]
\centerline{\includegraphics[width=15 cm, keepaspectratio]{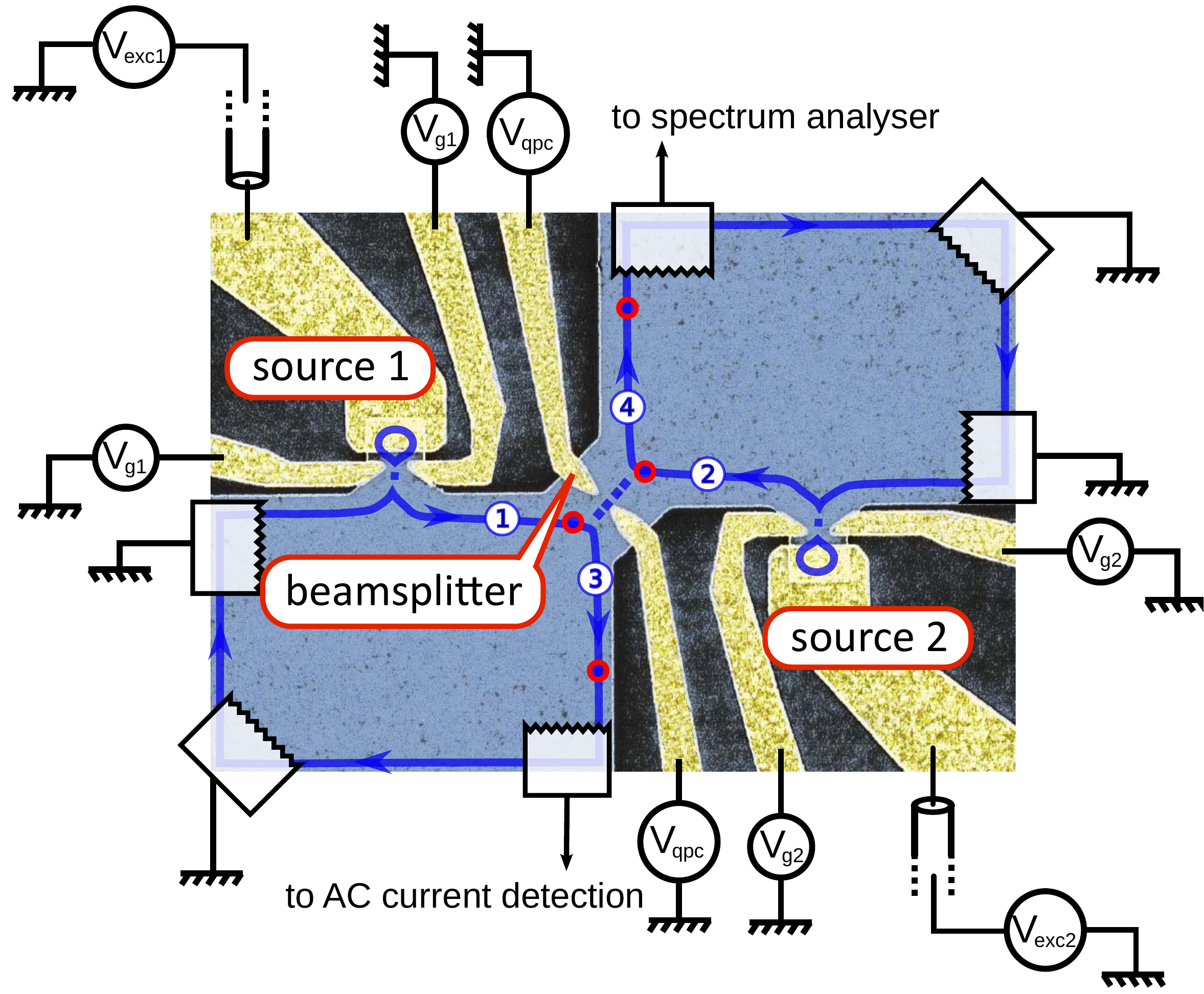}}
\caption{Sketch of the sample based on a SEM picture. The electron gas is represented in blue. Two single-electron emitters are located at inputs $1$ and $2$ of a quantum point contact used as a single electron beamsplitter. Transparencies $D_1$ and $D_2$ and static potentials of dots $1$ and $2$ are tuned by gate voltages $V_{g,1}$ and $V_{g,2}$. Electron/hole emissions are triggered by excitation drives $V_{exc,1}$ and $V_{exc,2}$. The transparency of the beamsplitter partitioning the inner edge channel (blue line) is tuned by gate voltage $V_{qpc}$ and set at $\mathcal{T}=1/2$. The average ac current generated by sources $1$ and $2$ are measured on output 3 while the low frequency output noise $S_{44}$ is measured on output $4$.
}
\end{figure}

\begin{figure}[hhhhhh]
\centerline{\includegraphics[width=15 cm, keepaspectratio]{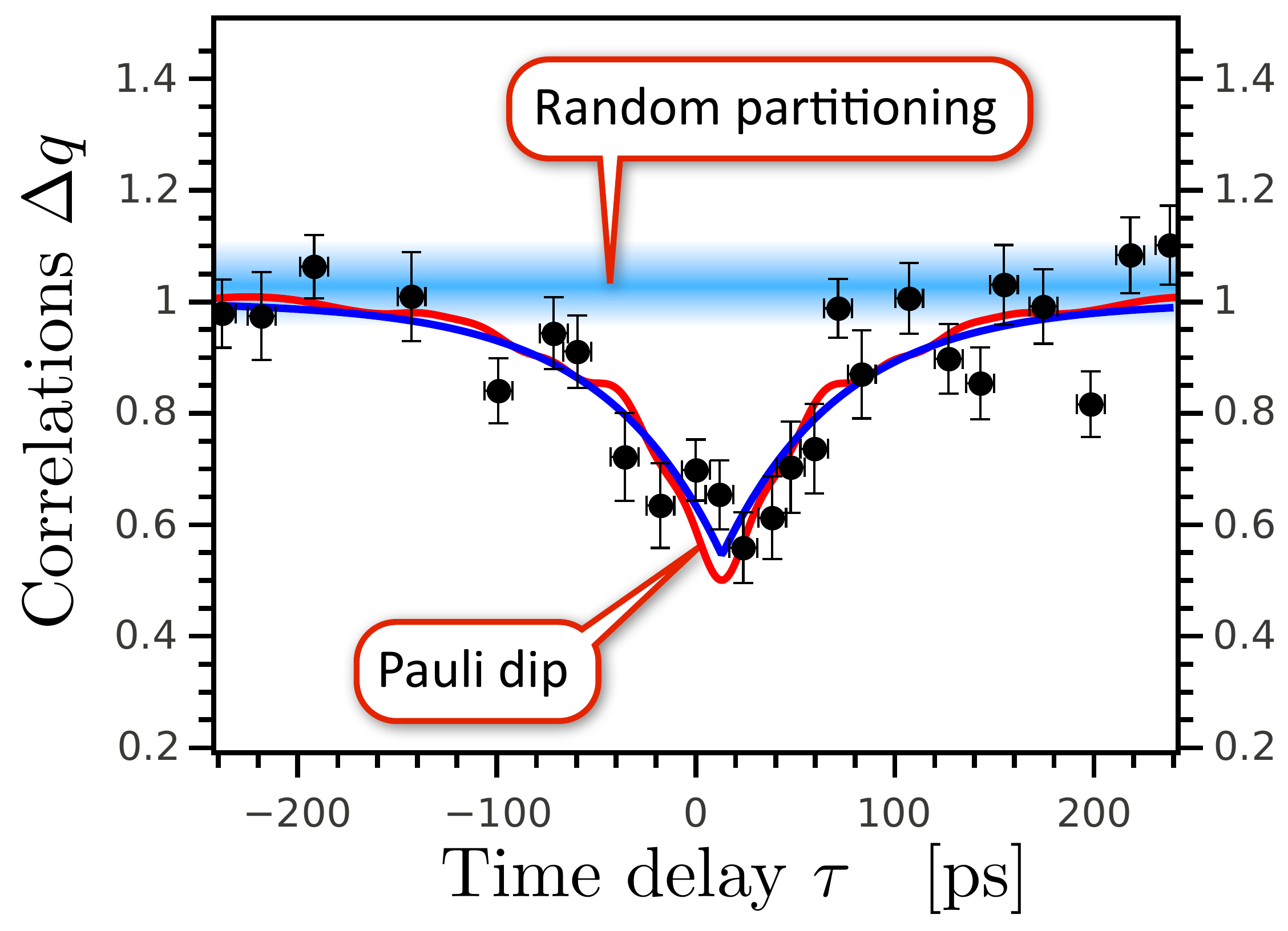}}
\caption {Excess noise $\Delta q$ as a function of the delay $\tau$ and normalized by the value on the plateau observed for long delays. The blurry blue line represents the sum of the partition noise of both sources.  The blue trace is an exponential fit by $\Delta q = 1- \gamma e^{- |t-\tau_0|/\tau_e}$. The red trace is obtained using Floquet scattering theory which includes finite temperature effects.}
\end{figure}

\end{document}